\documentstyle[aps,prb,overcite,twocolumn,psfig,rotating]{revtex}
\begin{document}
\draft

\title{Microscopic Theory of Protein Folding Rates.I: Fine Structure
of the Free Energy Profile and Folding Routes from a Variational Approach}

\wideabs{
\author{ John J. Portman \footnotemark[1],
Shoji Takada \footnotemark[2], and Peter G. Wolynes \footnotemark[1]\footnotemark[3]}
\address{\footnotemark[1] Departments of Physics and Chemistry, 
	University of Illinois, Urbana, Illinois, 61801\\  }
\address{\footnotemark[2] Department of Chemistry, 
Kobe University, Rokkodai, Kobe, 657 Japan\\}
\address{\footnotemark[3] Current address: Department of Chemistry and Biochemistry, 
University of California at San Diego, La Jolla, California 92093}

\date{August 23, 2000}
\maketitle

\begin{abstract}
{A microscopic theory of the free energy barriers and folding routes for minimally
frustrated proteins is presented, 
greatly expanding on the presentation of the 
variational approach outlined previously
[J. J. Portman, S. Takada, P. G. Wolynes, Phys.\ Rev.\ Lett. {\bf 81}, 5237 (1998)].
We choose the $\lambda$-repressor protein as an illustrative example and
focus on how the polymer chain statistics influence
free energy profiles and partially ordered ensembles of structures.  
In particular, we investigate the role of chain stiffness on the 
free energy profile and folding routes.  We evaluate
the applicability of simpler approximations in which the conformations of the
protein molecule along the folding route are restricted to have residues 
that are either entirely folded or unfolded in contiguous stretches.  We find
that the folding routes obtained from only one contiguous 
folded region corresponds to a 
chain with a much greater persistence length than appropriate for natural 
protein chains, while the folding route obtained 
from two contiguous folded regions is able to capture
the relatively folded regions 
calculated within the variational approach.
The free energy profiles obtained from the 
contiguous sequence approximations
have larger barriers than the more microscopic variational theory
which is understood as a consequence of partial ordering.
}
\end{abstract}
}
%************************************************************************
\section{Introduction}
%************************************************************************
Considerable progress has been made in describing protein folding
using equilibrium and nonequilibrium statistical mechanics, but 
a complete formal microscopic kinetic theory has only been sketched.  
The primary
novel features of the modern theory of folding revolve around two somewhat
different themes \cite{jdb:pgw:87,jdb:pgw:89,bosw:95} --- 
the glassy dynamics expected for most heteropolymers whose sequence 
is chosen at random 
\cite{jdb:pgw:89,bosw:95,dt:va:jkb:96,egt:yak:kad:96,st:jjp:pgw:97},
and the organized dynamics expected for proteins selected by evolution
to fold quickly on a funneled energy landscape 
\cite{pel:mm:jno:92,owls:95,hn:aeg:jno:98,js:jno:clb:99,ch:zls:pgw:99,vsp:dsr:99}.  
The establishment, through selection, of a funneled landscape
entails minimizing frustration \cite{jdb:pgw:87} 
--- the conflict between different energy contributions in a random sequence.  
Once a funneled landscape is established
by selection, however, it is the interplay between entropy and
guiding energies of the funnel that figure most prominently in 
determining the observed kinetics.  If we neglect sidechain and solvent
degrees of freedom, the entropy depends crucially on the polymer 
physics of the protein chain.  The statistical mechanical
theory of both the stable states and some transition states
has already been outlined \cite{ms:pgw:92,st:pgw:97a,st:pgw:97b,jjp:st:pgw:98}.  
In this paper and its companion \cite{jjp:st:pgw:00b}
we show how the existing framework can be extended to yield a complete
microscopic theory of folding rates of completely minimally
frustrated proteins.  
Microscopic calculations of transition state ensembles, activation free
energies, and dynamical pre-factors involving  chain motions
can be obtained.  A brief report on the early progress of this
work has already appeared \cite{jjp:st:pgw:98}, but here
we fill in the details and also explore some interesting polymer
physics issues that we only touched on previously.

The folding transition can be considered as a finite-size 
phase transition involving two or more stable phases: 
one a high entropy denatured state with little structure, although perhaps
collapsed; and the other the low entropy folded state \cite{privalov:79}.  
A quantitative way to distinguish among
these phases is through the magnitude of the fluctuations of 
each monomer about its average position.  
In the denatured states,
the protein explores many conformations, and because
there is little well defined structure, 
the fluctuations of a residue about any 
particular position are relatively large. In the folded state, 
the conformations are much more restricted and 
can be described as relatively small fluctuations about the 
localized positions of the average native
structure \cite{hf:gap:dt:79}.  
These small amplitude deviations are reflected
in part by the ``temperature factors'' (or Debye-Waller 
factors) which can be experimentally 
obtained from fitting X-ray crystallography data to a model 
structure that allows for these fluctuations.  

The qualitative difference between the liquid and solid phases in an 
ordinary first order crystallization transition can also 
be described by fluctuation magnitudes.  
The thermodynamics and to some extent the kinetics of 
first order phase transitions can be
studied using a free energy density functional \cite{oxtoby:91}. 
To analyze the kinetics of the transition one introduces
an approximate density profile that is able to interpolate
between the two phases, modeling the formation of crystallites or
droplets.  
The stable phases which are minima of the functional 
and the critical nucleus which is a saddle-point of the functional
can then be described through 
the variational parameters of the trial density.
From the point of view of structure, 
the folded state of the protein
is similar to one specific minimum of an amorphous solid
in so far as it is not infinitely periodic.
One major difference between most inorganic solids
and proteins is that proteins are polymers where 
topology is important; the residues not only have a chemical identity 
but also a definite sequence.  Accordingly, the uniqueness of 
the folded structure refers not only to the 3-dimensional shape, 
but also to the specific residues localized at the coordinates of the
native structure.  This suggests that in order to apply 
the free energy functional formalism
to study the folding of a particular protein, one should use order 
parameters that are local in sequence.

An approach to folding kinetics based on 
a ``site-resolved'' protein folding free energy functional was presented in 
Ref. \citen{jjp:st:pgw:98}.
In this approach, a variational free energy surface is introduced
directly through a reference Hamiltonian, which provides a good way of 
approximating the density.  
The resulting free energy surface is parameterized by the fluctuations of each 
residue about the average, folded conformation.
When put into practice the scheme is very similar in spirit to the
density functional calculations described above.  
The variational approach has
also been used to characterize the protein folding phase 
diagram \cite{ms:pgw:92,st:pgw:97a} 
as well as to study folding nucleation \cite{st:pgw:97b} without attending to
the specifics of a given native protein structure. 
In order to explore general issues of the role of the polymer chain
statistics in folding, 
we specialize our calculation 
to study a particular protein that folds to a known structure.  
The example we choose, the $\lambda$-repressor protein, has been much
studied in the laboratory \cite{bhdfo:96,bhdco:97a} 
but we will explore how its folding
routes and free energy profile are changed upon varying polymer statistics
sometimes using ``unphysical'' values of the backbone parameters 
in order to gain insight.

%********************************************************************
\section{Gaussian Model for a Stiff Chain}\label{ss:chain}
%********************************************************************
Consider the Gaussian approximation to the probability density for the 
$n$ monomer positions $\{{\bf r}_i\}$ of a polymer chain
\begin{equation}\label{eq:prob_r}
\Psi[\{ {\bf r}\}]
\sim \exp \left[ - \frac{3}{2 a^2} \sum_{ij}
 {\bf r}_i \cdot \Gamma_{ij} \cdot  {\bf r}_j \right],
\end{equation}
where $a$ is a microscopic length scale taken to be the mean square 
distance between adjacent monomers, and we have assumed the mean position
vanishes. 
The correlations of monomer positions in 
Eq.(\ref{eq:prob_r}) are given by 
$\langle  {\bf r}_i \cdot  {\bf r}_j \rangle / a^2 
= [\Gamma^{-1}]_{ij}$.

Different choices of the correlations result in different Gaussian 
models for the polymer
backbone.  It is most natural to define the model chain in terms of the 
correlations between the $n-1$ bond vectors:
${\bf a}_i =  {\bf r}_{i+1} -  {\bf r}_{i}$.
Denoting correlations between bond vectors by 
$\langle {\bf a}_i \cdot {\bf a}_j \rangle / a^2 
= [\Gamma^{({\rm b})-1}]_{ij}$,
the bond correlations and the monomer position correlations are related by
\begin{equation}\label{eq:bondcorr}
\Gamma = M^{T} \Gamma^{({\rm b})} M,
\end{equation}
where $M$ is the $(n-1) \times (n)$ nearest neighbor difference matrix
\begin{equation}\label{eq:M}
M = 
\left[
\begin{array}{ccccc}
-1     & 1     &       &\cdots & 0\\
       & -1    & 1     &       & \vdots \\
\vdots &       &\ddots & \ddots \\
0      &\cdots &       & -1    & 1 \\
\end{array}
\right]
\end{equation}

A simple way to account for chain stiffness is to assume
there is a fixed angle $\theta$ between
adjacent bonds. This freely rotating chain model \cite{flory:69} has 
monomer correlations that decay exponentially as
$\langle {\bf a}_i \cdot {\bf a}_{i+l} \rangle / a^2 = g^l$, where
$g = \cos \theta$.  Inverting the matrix of bond correlations and
using Eq(\ref{eq:bondcorr}) gives the inverse of the monomer correlations 
\cite{mb:rz:78}
\begin{equation}\label{eq:gamma}
\Gamma = \frac{1-g}{1+g} K^R 
+ \frac{g}{1-g^2}[K^R]^2 - \frac{g^2}{1-g^2} \Delta,
\end{equation}
where $K^R$ is the Rouse matrix for a nearest neighbor harmonic chain
\begin{equation}\label{eq:Krouse}
K^R = 
\left[
\begin{array}{ccccc}
1      & -1    &        &\cdots & 0 \\
-1     & 2     & -1     &       & \vdots \\
       &\ddots & \ddots & \ddots\\
\vdots &       & -1     & 2     & -1 \\
0      &\cdots &        & -1    & 1 
\end{array}
\right],
\end{equation}
and $\Delta$ is accounts for the ``boundaries'' at the
end of the chain
\begin{equation}\label{eq:delta}
\Delta = 
\left[
\begin{array}{cccccc}
1      & -1    & \vline &       &\cdots & 0 \\
-1     &  1    & \vline &       &       &\vdots\\
\cline{1-3}\\
\cline{4-6}
\vdots &       &        &\vline & 1     & -1 \\
0      &\cdots &        &\vline & -1    &  1
\end{array}
\right].
\end{equation}

In the continuum limit, this Gaussian polymer model 
gives the familiar form of wormlike chain that restricts the mean 
separation between adjacent monomers as well as the local 
curvature of the chain 
(e.g., see Ref. \citen{dt:byh:98} and references therein).
In this paper, we use the 
discrete representation and identify the monomers to be the 
$\alpha$ carbons composing the polypeptide backbone.  With this choice,
the root mean square separation distance $a$ is the typical distance between 
adjacent $\alpha$ carbons $a \approx 3.8$\AA.

The other parameter in the chain model
defines the chain stiffness: as
$g \rightarrow 0$, $\Gamma = K^R$ gives the familiar correlations
of a flexible chain 
($\langle {\bf a}_i \cdot {\bf a}_j \rangle = a^2 \delta _{ij}$), 
and as $g \rightarrow 1$, the correlations correspond
to that of a rigid rod 
($\langle {\bf a}_i \cdot {\bf a}_j \rangle = a^2$).  
Another measure of the chain stiffness is given by the 
persistence length 
$l_i = \sum_j \langle {\bf a}_i \cdot {\bf a}_j \rangle /a $ 
\cite{flory:69}.  It is possible to introduce non-uniform 
stiffness parameters (and hence local persistence lengths)
in order to model the different flexibilities 
of the monomers composing a heteropolymer \cite{perico:89b}.  
For example, the bond correlations can be extracted
from the rotational isomeric states model of Flory \cite{flory:69},
or from equilibrium simulations of a detailed polymeric potential
\cite{perico:89b,ap:fg:ga:87,perico:89,hffp:91,ap:nem:mde:98}.  
Still more complex models can be used to model 
explicit chiral helical tendency \cite{yamakawa:97}
through anisotropic Gaussian correlations.
For simplicity, we assume here that the
chain stiffness is uniform in this paper, so that the persistence length
is related to the chain stiffness by $l \approx a/(1-g)$.  
For proteins, a
reasonable value for the chain stiffness is $g = 0.8$ which corresponds
to the persistence length of polyalanine, 
$l = 5 a \approx 20$\AA  \cite{flory:69,cantor_schimmel:80}.  
In this paper we will also compare chains of other uniform 
stiffnesses as well.

In addition to the Gaussian correlations given by 
Eq.(\ref{eq:prob_r}) and Eq.(\ref{eq:gamma}) for a stiff chain,
we also include a confining potential that controls the overall size
of the polymer chain, which is to say it determines the proximity to 
the chain collapse transition.
To model a collapsed stiff chain, we consider the 
chain Hamiltonian
\begin{equation}\label{eq:hchain1}
\beta H_{{\rm chain}} 
= \frac{3}{2a^2} 
\sum_{ij} {\bf r}_i \cdot \Gamma_{ij} \cdot {\bf r}_j + 
\frac{3}{2a^2} B \sum_i {\bf r}_i^2.
\end{equation}
The second term controls the degree of collapse of the chain
through a confining potential where the parameter $B$ is 
conjugate to the radius of gyration of the chain.  To establish
notation for future use, we rewrite Eq.(\ref{eq:hchain1}) as
\begin{equation}\label{eq:hchain}
\beta H_{{\rm chain}} 
= \frac{3}{2a^2} 
\sum_{ij} {\bf r}_i \cdot [\Gamma^{{\rm(ch)}}]_{ij} \cdot {\bf r}_j,
\end{equation}
where 
\begin{equation}\label{eq:gamma_ch}
\Gamma^{{\rm (ch)}}_{ij} = \Gamma_{ij} + B\delta_{ij}
\end{equation}
are the inverse monomer correlations of the collapsed stiff chain.

%*******************************************************************
\section{Variational Free Energy Surface}
%*******************************************************************

\begin{figure}%[h]
\psfig{file=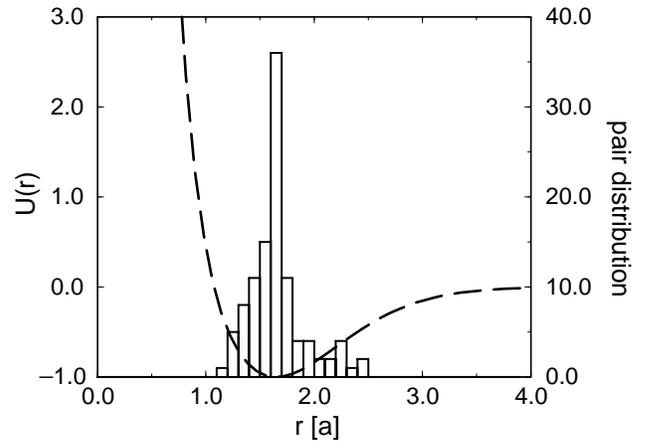,width=3.3in}
\caption{Interaction potential and $\lambda$-repressor native contact distribution.
The intermediate- and long-range parameters are given in the text.}
\label{fig:u(r)}
\end{figure}

The Hamiltonian for our protein model is 
\begin{equation}\label{eq:H}
H = H_{{\rm chain}} + H_{{\rm int}},
\end{equation}
where $H_{{\rm chain}}$ is the backbone potential defining the polymeric 
correlations given in Eq.(\ref{eq:hchain}) and 
$H_{{\rm int}}$ is the (2-body) interaction 
potential between distant monomers.  
The interaction between distant monomers are modeled by 
a pair potential $u(r)$
\begin{equation}\label{eq:hint}
H_{{\rm int}} = \sum_{[ij]} \epsilon_{ij} u(|{\bf r}_i - {\bf r}_{j}|),
\end{equation}
where $\epsilon_{ij}$, the strength of the interaction, depends on the
identity of the residues $i$ and $j$.  
The spatial dependence of the interactions between distant monomers 
consists of an attractive well and a repulsive core.
For computational convenience, we approximate the 
interaction potential as the sum of three Gaussians:
\begin{equation}\label{eq:uint}
u(r) = \sum_{k = ({\rm s,i,l})} \gamma_j 
\exp \left[ - \frac{3}{2a^2} \alpha_k r^2 \right].
\end{equation}
where
$(\alpha_{{\rm s}} > \alpha_{{\rm i}} > \alpha_{{\rm l}})$
are the ranges of the short-, intermediate-, and long-range
interactions, respectively.  The intermediate-range term is repulsive 
$(\gamma_{{\rm i}} > 0)$
and the long-range term is attractive $(\gamma_{{\rm l}} < 0)$; 
the intermediate- and long-ranged potential parameters 
are chosen so that the sum
of these two terms gives a potential well at an appropriate distance
for contacts in the native structure.
The short-range term is repulsive $(\gamma_{{\rm s}} > 0)$
and represents the hard core repulsion between residues. 
(For an example of $u(r)$ see Fig. \ref{fig:u(r)})

We approximate the free energy surface of the protein using a reference
Hamiltonian that corresponds to a polymer in a non-uniform external field that
constrains the monomers to lie near their locations in the native structure 
$\{{\bf r}_i^{{\rm N}}\}$
\begin{equation}\label{eq:h0}
\beta H_0  =  \beta H_{{\rm chain}} + 
\frac{3}{2a^2} \sum C_i ({\bf r}_i - {\bf r}_i^{{\rm N}})^2 .
\end{equation}
The strengths of the harmonic constraints, $\{C_i\}$, are conjugate to
the fluctuations of the polymer about each of the native positions.  
The external constraints in $H_0$ influence both the 
correlations $G_{ij}$ and average positions $\{\mathbf{s}_i\}$ 
of the monomers composing the reference chain:
\begin{equation}\label{eq:G}
G_{ij} \equiv \langle \delta {\bf r}_i \cdot \delta {\bf r}_j \rangle_0/a^2
= [\Gamma^{(0)}]^{-1}_{ij}
\end{equation}
\begin{equation}\label{eq:rave}
{\bf s}_i \equiv 
\langle {\bf r}_i \rangle_0 =  \sum_j G_{ij} C_j {\bf r}_j^{{N}},
\end{equation}
where $\delta {\bf r}_i$ is the position of the $i^{{\rm th}}$ monomer
relative to the average
\begin{equation}\label{eq:deltar}
\delta {\bf r}_i = {\bf r}_i - \langle {\bf r}_i \rangle_0 
={\bf r}_i - {\bf s}_i,
\end{equation}
and
$\Gamma^{{\rm (0)}}$ is the matrix of coefficients of the quadratic terms 
of $H_0$
\begin{equation}\label{eq:gam0}
\Gamma^{{\rm (0)}}_{ij} = \Gamma^{{\rm(ch)}}_{ij} + C_i \delta_{ij}.
\end{equation}
From the magnitude of these fluctuations, 
this reference 
Hamiltonian can distinguish the two stable phases of the protein (as described above): 
the globule corresponds to large fluctuations (weak constraints) 
and folded states correspond small fluctuations (strong constraints).

We consider the variational free energy surface parameterized by the 
constraint parameters $\{C_i\}$
\begin{equation}\label{eq:fvar_def}
F[\{C\}] = - k_B T \log Z_0 + \langle H - H_0 \rangle_0,
\end{equation}
where $Z_0 = {\rm Tr}\left[e^{-\beta H_0} \right]$ is the partition
function of the reference Hamiltonian, and 
$\langle \ldots \rangle_0 = {\rm Tr}\left[ \ldots e^{-\beta H_0}\right]/Z_0$ 
denotes the average taken with respect to $H_0$.  Substituting the 
expressions for $H$ and $H_0$ gives the variational free energy 
$F = E - ST$, where
\begin{equation}\label{eq:evar}
E[\{C\}] = \sum_{[ij]} \epsilon_{ij} 
\langle u(|{\bf r}_i - {\bf r}_j|)\rangle_0
\end{equation}
and
\begin{equation}\label{eq:svar}
S[\{C\}]/k_B = \log Z_0 + 
\frac{3}{2a^2} \sum C_i \left\langle 
({\bf r}_i - {\bf r}_i^{{\rm N}})^2
\right\rangle_0
\end{equation}
are the expressions for the energy 
$E[\{C\}]$ and entropy $S[\{C\}]$ as functions of the 
variational constraints. 

The averages over $H_0$ in Eq.(\ref{eq:evar}) and Eq.(\ref{eq:svar})
can be expressed in terms of $G$ and $\{{\bf s}_i\}$,
because $e^{-\beta H_0}$ is a Gaussian distribution. 
One instructive way to calculate the averages 
is to introduce approximations to the 
density of monomer $i$, 
$\rho_i^{1}({\bf r}) = \langle \delta({\bf r} - {\bf r}_i) \rangle_0$, 
\begin{equation}\label{eq:dens0}
\rho_i^{1}({\bf r}) = \left[ \frac{3}{2 \pi a^2 G_{ii}}\right]^{3/2}
\exp \left[-\frac{3({\bf r} - {\bf s}_i)^2}{2 a^2 G_{ii}} \right],
\end{equation}
and the pair density between $i$ and $j$,
$\rho_{ij}^{2}({\bf r}) = 
\langle \delta \left( {\bf r} - ({\bf r}_{i} -{\bf r}_{j})\right) \rangle_0$,
\begin{equation}\label{eq:pdens0}
\rho_{ij}^{2}({\bf r}) = 
\left[ \frac{3}{2 \pi a^2 \delta G_{ij}} \right]^{3/2}
\exp \left[ 
-\frac{3({\bf r} - ({\bf s}_{i} - {\bf s}_{j}))^2}{2 a^2 \delta G_{ij}}
\right],
\end{equation}
where  
$\delta G_{ij} = 
\langle (\delta {\bf r}_i - \delta {\bf r}_j)^2 \rangle_0/a^2
= G_{ii} + G_{jj} - 2 G_{ij}$.  
These densities depend on the constraint parameters $\{C_i\}$ 
through $G$ and $\{{\bf s}_i\}$.
Averages over $H_0$ can be calculated through 
$\rho_i^{1}({\bf r}) $ and $\rho_{ij}^{2}({\bf r}) $, for example,
\begin{equation}\label{eq:exave}
\langle u(|{\bf r}_i-{\bf r}_j|) \rangle_0 = 
\int \! \! d{\bf r} \:\rho_{ij}^{2}({\bf r}) u(r).
\end{equation}
In this way, the variational free energy can be viewed as a density 
functional with a particular approximation to the density that 
simultaneously incorporates the polymeric correlations and 
the monomeric fluctuations about the average positions.

It is straightforward to calculate the entropy
and energy in terms of the monomer correlations and mean positions.
After some manipulation (see Appendix), the entropy can be written as 
\begin{equation}\label{eq:svar2}
S[\{C\}] = \frac{3}{2} \log \det G - 
\frac{3}{2a^2}\sum_{ij} s_i \cdot \Gamma^{{\rm (ch)}}_{ij} \cdot s_j 
+ \frac{3}{2}\sum_i C_i G_{ii}.
\end{equation}
This expression can be interpreted as follows:
the first term is the entropy of the chain due to polymeric 
fluctuations, the second term is the entropy loss of fixing 
each monomer to the average positions, and the last term 
is the entropy of the vibrations about the mean position
$(= (3/2a^2)\sum C_i \langle \delta {\bf r}_i^2 \rangle_0)$.  Similarly,
the pair potential can be averaged over $H_0$ to give the energy
\begin{equation}\label{eq:evar2}
E[\{C\}] = \sum_{[ij]} \epsilon_{ij} u_{ij},
\end{equation}
where
\begin{eqnarray}\label{eq:uave}
u_{ij} & = &\langle u(|{\bf r}_{ij}|) \rangle_0 \\
       & = &\sum_{k = ({\rm s,i,l})} 
\frac{\gamma_k}{(1 + \alpha_k \delta G_{ij})^{3/2}}
\exp 
\left[-\frac{3}{2a^2}
\frac{\alpha_k ({\bf s}_i - {\bf s_j})^2}{1+\alpha_k \delta G_{ij}}
\right]. \nonumber
\end{eqnarray}
Finally, we choose to measure the free energy relative to the unconstrained
chain
\begin{equation}
\Delta F[\{C\}] = \Delta E[\{C\}] - T \Delta S[\{C\}],
\end{equation}
where, for example, $\Delta F[\{C\}] = F[\{C\}] - F[\{C = 0\}]$.

The reference Hamiltonian plays such a prominent role in the 
variational theory (and in subsequent calculations of folding dynamics) that
it warrants further comment.  
Not surprisingly, other Gaussian models have been previously introduced 
to model the polymers with fixed contacts or crosslinks.
A Hamiltonian of the form
\begin{equation}\label{eq:hcon}
\beta H_{{\rm h.c.}} = \frac{3}{2a^2} \sum ({\bf r}_i - {\bf r}_{i+1})^2 + 
C \sum_{[ij]} ({\bf r}_i - {\bf r}_j)^2
\end{equation}
has been used to study the thermodynamics 
\cite{flory:76,mps:tav:95,jdb:dt:96,mps:tav:96,zwanzig:96,vilgis:98}
and dynamics \cite{mps:tav:97} of polymers crosslinked at the sites
specified by the pairs $[ij]$.
This harmonic contact Hamiltonian, $H_{\mathrm{h.c.}}$,
has also been used to model the vibrations of 
folded proteins \cite{ib:ara:be:97,bade:98,ib:aw:dbc:98},
with the set $[ij]$ limited to the contacts of the native structure, and 
it was found that the relative magnitudes of monomer 
fluctuations agree well with measured temperature 
factors \cite{ib:ara:be:97}.  
In contrast to our reference Hamiltonian, $H_{\mathrm{h.c.}}$
is translationally invariant, independent of an explicit native
structure.
While enforcing this symmetry has some advantages, the 
potential well is centered at the origin.
We note that recently, $H_{\mathrm{h.c.}}$ was used as a 
mean spherical model
for protein folding where a nonlinear constraint was added so that
the average monomer positions would
lie on the surface of a sphere, preventing the
polymer conformation from collapsing to the origin; the minima of 
$H_{\mathrm{h.c.}}$ with the added imposed condition can yield
a meaningful average structure \cite{be:kad:00}.
Nevertheless, the mean locations of the residues are quite 
distorted.
Consequently, $H_{\mathrm{h.c.}}$ 
is not as well suited as the Hamiltonian we use
to describe the protein folding transition where 
the disordered globule and the structured folded state are 
separated by a barrier composed of an ensemble of partially ordered 
conformations.
We have, however, investigated $H_{\mathrm{h.c.}}$ as a reference
Hamiltonian in the variational context. 
It gives similar results to those described here.

%-------------------------------------------------------------------------
\section{Order Parameters and Folding Routes}
%-------------------------------------------------------------------------

Setting values for the constraint parameters $\{C\}$ corresponds to selecting an
ensemble of conformations specified by $\{{\bf s}\}$ and $G$.
The energy of a pair is most stabilizing when the pair density is contained 
in the potential well, i.e., the mean separation between monomers 
is within the well and the fluctuations
are relatively small.  Accompanying this stabilization, however, 
is the entropy loss of localizing the positions of the pair. 
In general, when there are many non-zero constraints, the entropy loss
due to localization is given by Eq.(\ref{eq:svar2}) through the correlations.  
In other free energy functionals 
\cite{bas:jw:pgw:97,bas:jw:pgw:99,ovg:avf:99,ea:db:99,vm:wae:99}
this entropy loss is estimated
but there are difficulties in considering only the entropy loss
of individual pairs forming loops because the total entropy loss is 
inherently nonadditive \cite{hsc:kad:89,hsc:kad:90}.
The relation of the two approaches is much like the difference
between the Thomas-Fermi and Hohenberg-Kohn estimates of the kinetic
energy in quantum mechanical density functional theory \cite{parr:yang:89}.
The values of the constraint parameters corresponding to the local minima of
$F[\{C_i\}]$ are a compromise between the energy and entropy decrease
of forming contacts.
Similarly, the constraints that correspond to 
saddle-point of $F[\{C_i\}]$ also reflect this competition of energy and entropy,
because the free energy is minimized in all directions except the unstable 
mode along which there is a maximum.

The configurational ensemble corresponding to a 
given set of constraint parameters $\{C_i\}$ can also be described 
by density like order parameters that depend on the local mean square
fluctuations.
For any set of functions of the chain positions
$\{A_i[\{{\bf r}\}]\}$,
we can define the order parameter 
$A_i[\{C\}] = \langle A_i \rangle_0$, as a function of the constraints.
This relationship can be inverted locally provided the 
Jacobian is nonsingular, 
$\det J \neq 0$ with $J_{ij} = \partial A_i/\partial C_j$.
Since these order parameters are a function of the constraints 
we can then parameterize the free energy by 
$\widetilde{F}[\{{\cal A}\}] = F[\{C\}]$ 
with 
$C_i = C_i[\{A\}]$.
For example, the form of $H_0$ suggests that the
local mean square fluctuations (related to Debye-Waller factors)
\begin{equation}\label{eq:tfac}
{\cal B}_i = \langle \delta {\bf r}_i^2 \rangle_0 = G_{ii} a^2
\end{equation}
are a natural set order parameters for the reference Hamiltonian.
(Indeed, this is what motivated our choice of $H_0$).
In studying the dynamics of barrier crossing in the companion
paper, it will prove useful to consider a related but 
different measure of native similarity. 
With any identification of the order parameters, we can study
the properties of the free energy in $\{C_i\}$-space to describe the 
folding, and then characterize the corresponding ensembles through
structural order parameters as equilibrium 
averages with Hamiltonian $H_0$.

We calculate the transition states involved in the folding
by searching for saddle-points in $F[\{C_i\}]$ using an 
eigenvector-following algorithm \cite{wales:94}.  
This algorithm is similar to Newton's 
method for optimization, but involves diagonalizing the Hessian matrix, 
$\partial^2 F/\partial C_i\partial C_j$, at each iteration.  
In this routine, the point is updated by stepping in a direction 
to maximize along the eigenvector with the lowest eigenvalue and 
minimize along along all others.  To find a minimum, a step is taken 
to minimize along all eigenvectors of the Hessian. 
In order to use this algorithm, we need to be able to differentiate the
free energy with respect to $\{C_i\}$,
$\partial_{\alpha}F \equiv \partial F/\partial C_{\alpha}$.  
These derivatives can be easily computed by the chain rule using
the elementary derivatives
$\partial_{\alpha} G_{ij} = - G_{i\alpha}G_{\alpha j}$ and 
$\partial_{\alpha}(\log \det G) = - G_{\alpha \alpha}$.
The explicit expressions for the derivatives of the energy and entropy
are not given here; they are straight-forward to derive,
and not very illuminating.

The saddle-points and local minima characterize the average folding routes
from this theory.  These average pathways are found as follows.
The globule and native states are identified by the local minima with 
the largest and smallest entropy, respectively.  These are easy to 
identify, because the globule is the only stable minimum at high temperature
and the native is the only one at low temperature; these minima can be
used as the initial guesses for the optimization algorithm for
incremental temperature changes until we have these minima at the 
same temperature.
Using linear combinations of these two sets of constraints as 
initial guesses, we search for a saddle-point.  
From this saddle-point, we perturb the set of constraints $\{C_i\}$ 
along the unstable 
eigenvector and use the eigenvector following algorithm with a small
step size to find the closest minimum.  This gives two local minima, 
one for each
direction on the unstable eigenvector, connected by the saddle-point.
This process is repeated until the globule and native state are
connected by a series of local minima and saddle-points.
We identify this connected sequence as the average folding route, 
characterizing the transition states and local minima that are
important in the folding kinetics.
We note that in the example below for the $\lambda$-repressor protein,
only one folding route was found, but this is not a general result of
the theory.  The same procedure applied to the SH3 domain
has shown that there may be multiple routes (unpublished).

%************************************************************************
\section{Folding Routes Example: $\lambda$-repressor}
%************************************************************************

In this section, we illustrate the variational theory by studying the
folding of a variant of the $\lambda$-repressor protein.  
$\lambda_{6-85}$, is a small (80 residue) protein 
consisting of five helices in the native structure%
\cite{fn:ptw00a_lam}
\cite{1lmbpdb}.
 This protein folds extremely rapidly following two-state 
kinetics \cite{bhdfo:96}.  From NMR measurements of the folding 
rate of various mutants, Oas and coworkers concluded that the structure 
of the transition state consisted mainly of residues 
in helices H1 and H4 \cite{bhdco:97a}.
In Ref. \citen{jjp:st:pgw:98} we compared folding routes calculated from the
variational theory with these measurements,
using a reasonable choice for the persistence length of the polypeptide 
chain.  
We investigate here how these results depend on different values of the 
persistence length (or chain stiffness).  These studies allow us to see how some 
recent simplified approaches to free energy profiles based on assuming complete
contiguous sequence folding \cite{ovg:avf:99,ea:db:99,vm:wae:99}
become more exact as chain stiffness increases.

%=============================================================================
\subsection{Model Parameters}
%=============================================================================
To apply the theory, we need to specify the parameters that describe the 
interaction potential between residues and the polymer chain characteristics.  
The parameters of the present 
paper are the same as those chosen in Ref. \citen{jjp:st:pgw:98}, 
but we describe our rationale for this choice in greater detail.
For this fast folding protein, we consider a G\={o} model for the interactions.
This means that the sum over residues $[ij]$ in Eq.(\ref{eq:hint}) 
is limited to the set of contacts found in the native structure.
This set is defined to be pairs of residues $(i + 4 \le j)$ that have
$\beta$ carbons ($\alpha$ carbons for glycine) distances within a 
$6.5$\AA\ cutoff 
in the folded structure. A cutoff between $6 - 8$\AA\ 
is commonly used to define contacts in a G\=o model, though 
the precise value is not generally important.  We also include in
this set residue pairs that are likely to have hydrogen bonds 
(as determined by the DSSP algorithm \cite{wk:cs:83}) but fall outside 
the cutoff.  The strength of the interactions for this set depends on the 
residue identities of the pair. We take the well depths $\epsilon_{ij}$ 
to be the magnitude of the Miyazawa-Jernigan energy parameters reported 
in Ref. \citen{sm:rlj:96} in units of $\epsilon_0 = k_B T$.

The parameters for the interaction potential $u(r)$ in Eq.(\ref{eq:uint}) 
are constrained by the $C_\alpha - C_\alpha$ distances of the set of 
native contacts.
The intermediate- and long-range parameters are chosen so that 
the sum of these Gaussians has an attractive well that contains all the 
native contacts and has a minimum value $u(r^{\star}) = -1$ at the most 
probable contact distance $r^{\star} = 1.6 a$.  The contact distribution
and potential with 
$(\gamma_{\rm i},\alpha_{\rm i};\gamma_{\rm l},\alpha_{\rm l})$ = 
(9,0.54;-6,0.27) are shown in Fig. \ref{fig:u(r)}.  

\begin{figure}%[h]
\psfig{file=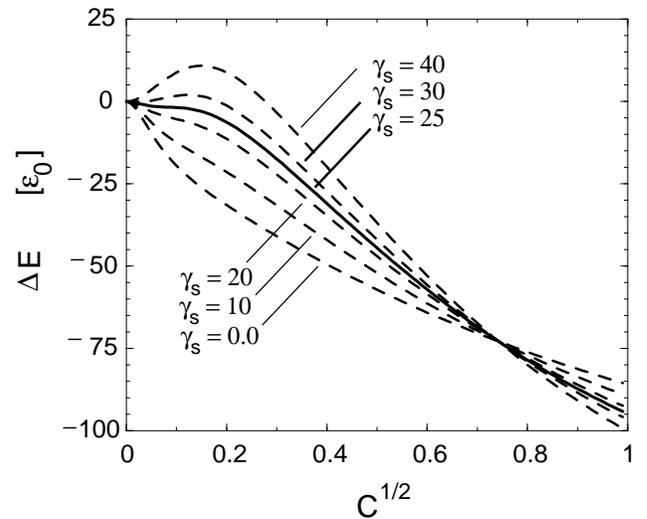,width=3.3in}
\caption{
Energy in units of the
Miyazawa-Jernigan energy scale $\epsilon_0$ 
as a function of the constraint, $C$. 
All monomers have equal constraint (1-dimensional), and 
chain stiffness $g = 0.8$. 
$\gamma_s$ is the strength of the short-range interaction
and the Gaussian width is $\alpha_s = 3$.  The intermediate- and
long-range interaction parameters are the same as in Fig. \ref{fig:u(r)}.
}
\label{fig:1de}
\end{figure}

The short-range 
interactions represent the hard core repulsion between the monomers 
which controls the density of the collapsed polymer.  Due to the Gaussian 
chain approximation, the pair density given by Eq.(\ref{eq:pdens0}) has 
non-zero density at short distances, and hence even
a de-localized pair has energy contributions from both the repulsive 
and attractive components of the potential.
In the model, the short-range Gaussian  amounts to an effective 
potential that balances the attractive potential for relatively 
unconstrained polymers.  
Choosing the repulsion by this criterion is analogous to finding
$\Theta$ solvent conditions for the unconstrained polymer 
(such as the globule).
To determine a reasonable repulsive potential for this model, we consider
a one-dimensional approximation to the variational free energy 
by setting all the constraints equal $C_i = C$.
As can be seen in Fig. \ref{fig:1de}, the energy as a function of $C$
is monotonically stabilizing if $\gamma_s$ is small, and has a barrier
if $\gamma_s$ large.  The parameters for the repulsive potential
are chosen so that the energy is relatively constant for small values
of the constraint parameter: 
$(\gamma_{\rm s}, \alpha_{\rm s}) =$ (25,3.0).  
(This particular value of the strength depends on the width of the 
short-ranged Gaussian which has chosen somewhat arbitrarily).

\begin{figure}%[h]
\psfig{file=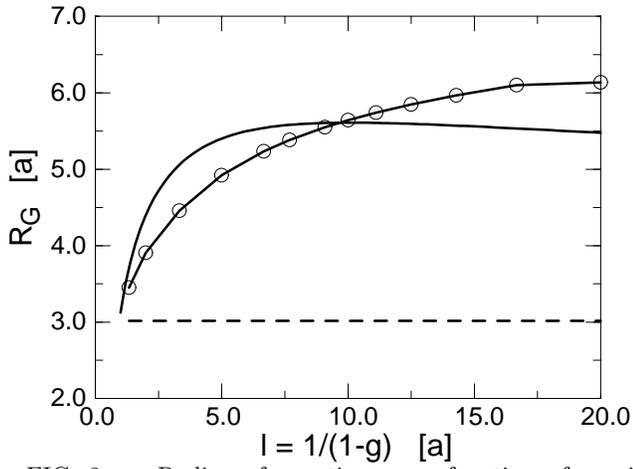,width=3.3in}
\caption{
Radius of gyration as a function of persistence length (in units of
monomer spacing $a$) for 
$\lambda$-repressor  $(n = 80)$ and confinement parameter $B=10^{-3}$: 
no constraints (solid), globule ($\bigcirc$), and native coordinates (dashed).}
\label{fig:rg_l}
\end{figure}

The remaining parameter to be specified is $B$, the strength of the
diagonal confinement term. This confinement parameter is a small 
constant that effects the fluctuations of unconstrained segments of
the chain since the constraint parameters are also diagonal terms in the 
inverse correlation matrix.  
Fig. \ref{fig:rg_l} shows the radius of gyration, 
\begin{equation}
R_G^2 = 1/n^2\sum_{i\le j} \langle({\bf r}_i - {\bf r}_j)^2\rangle_0,
\end{equation}
evaluated at the native coordinates (dashed) and an 
unconstrained chain, $C_i = 0$,
(solid) as a function of persistence length for $B = 10^{-3}$.  
The unconstrained radius of gyration rises rapidly as the persistence
length increases and saturates to a value less than twice the native
radius of gyration.  
The radius of gyration, which in the absence of confinement approaches
a large value in the stiff chain limit%
\cite{fn:ptw00a_rg}
, is seen to be limited by the confinement term.
This tension between
local stiffness and confinement is responsible for the plateau in $R_G$.
Fig. \ref{fig:rg_l} also shows the 
the radius of gyration evaluated at the constraints corresponding
to the globule minimum ($\bigcirc$).
For persistence lengths less than $l \approx 10a$ the globule constraints
lead to a smaller radius of gyration than the free chain value, and
for larger persistence length chains the radius of gyrations is somewhat
larger.  Although
it is possible to choose values of $B$ in order to set $R_G$ 
of the globule for each chain stiffness, we have chosen to illustrate
the effects on the polymer conformations by fixing the 
confinement and independently varying the chain stiffness.

%=========================================================================
\subsection{Two-dimensional Illustration}
%=========================================================================
Because the results of calculations in the full variational space 
(one constraint parameter for each monomer) are somewhat complicated to present, 
we will illustrate the model in lower dimensions to give the
reader an 
intuitive feeling for the multi-dimensional
free energy surface and how folding routes are
obtained from the model.
We consider a two-dimensional approximation in which we group
the protein into two segments and assign a variational 
spring constant to each group.  In this example, monomers with index $1-50$
correspond to $C_1$ and the rest correspond to $C_2$;
These two groups correspond roughly to the helices H1--H3 and H4--H5, respectively.
This grouping has a loose correspondence to the folding route obtained
from the full variational calculations discussed in the next section.

\begin{figure}%[h]
\psfig{file=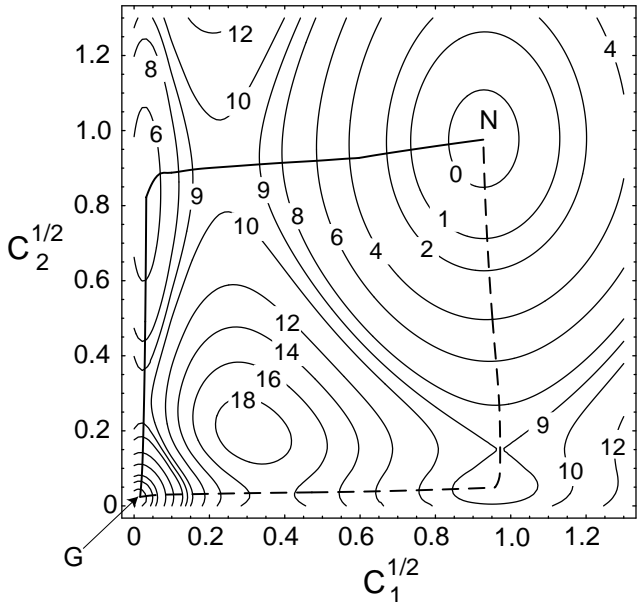,width=3.3in}
\caption{Contour plot of free energy  in units of the
Miyazawa-Jernigan energy scale $\epsilon_0$ for 
the two-dimensional surface (see text). The lines indicate
the average folding routes: Path 1 (dotted), Path 2 (solid).
The chain stiffness is $g = 0.8$}
\label{fig:2dcontour}
\end{figure}

The free energy surface shown in Fig. \ref{fig:2dcontour}
has two distinct low energy paths determined 
by the saddle-points connecting the globule (G) and native (N) states.
The average folding routes, as defined by the path
from the saddle points to the local minima, are determined 
by the eigenvector-following
algorithm.  As can be seen in Fig. \ref{fig:2dcontour} these routes
are very close to the steepest descents path.
Along the two routes from G to N, in Path 1 (dotted line)
the constraint $C_1$ progressively increase
followed by the increase of constraint $C_2$, whereas along path 
2 (solid line) this order is reversed with $C_2$ increasing before $C_1$.

\begin{figure}%[h]
\psfig{file=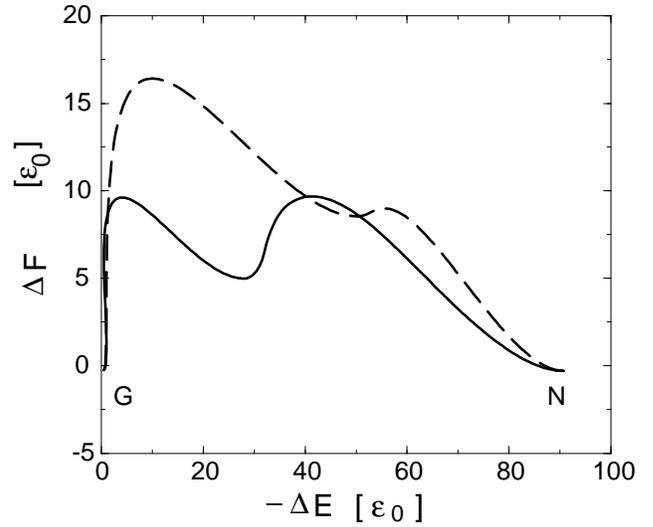,width=3.3in}
\caption{Free energy profile for both paths indicated in 
Fig. \ref{fig:2dcontour}.}
\label{fig:2d_fprofile}
\end{figure}

The free energy of these paths is plotted parametrically 
versus the energy ($\Delta F(C_1,C_2)$ vs. $\Delta E(C_1,C_2)$) in 
Fig. \ref{fig:2d_fprofile}, giving a free energy profile where 
the saddle-points appear as local maxima.  Path 2 is the 
relatively favored route since it has the lower barrier 
to folding.  We note that a rough description of the free energy 
profile can be represented by connecting the stationary points of 
$F(C_1,C_2)$ by straight lines.

\begin{figure}%[h]
\psfig{file=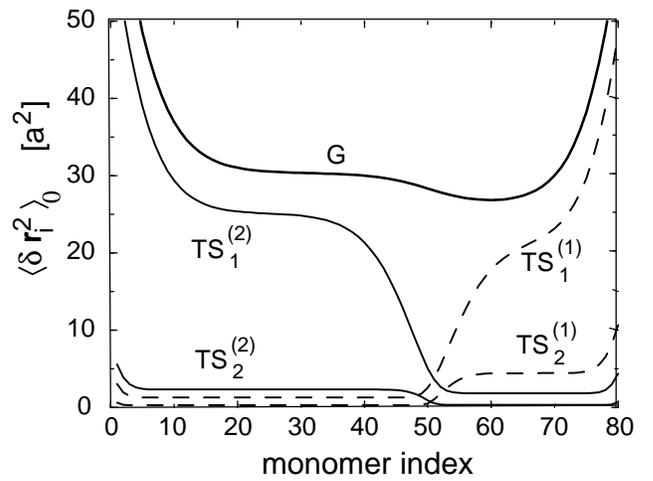,width=3.3in}
\caption{Fluctuations vs. sequence index,
$\langle \delta {\bf r}^2_i \rangle_0 = G_{ii} a$ 
where a is the distance between successive monomers, 
evaluated at the saddle-points for 
Path 1 (dotted) and Path 2 (solid) shown in Fig. \ref{fig:2d_fprofile}.  
Fluctuations of the Native (N) and Globule (G) are also shown.
}
\label{fig:2d_gii}
\end{figure}

The ensemble of conformations composing the average folding route can be 
characterized by the magnitude of the fluctuations of each residue
$({\mathcal B}_i = \langle \delta {\bf r}_i^2 \rangle_0)$ 
for any set of constraints along these paths.  
Fig. \ref{fig:2d_gii} shows the 
monomer fluctuations evaluated at the constraints corresponding to the
globule state and the saddle-points of both paths.  These fluctuations
give a description of folding consistent with 
the two-dimensional surface parameterized by the constraints: 
in path 1 residues in helices H1--H3 become structured at ${\rm TS}_1$ 
followed by H4--H5 at ${\rm TS}_2$, and in path 2 the order is reversed.  
For a given set of the constraints (i.e., a given saddle-point),
the fluctuations are seen to smoothly interpolate between the two groups of monomers.
The precise shape of the interface depends on the value of the chain 
stiffness.  A stiffer chain would tend to suppress variations as a function
of sequence index resulting in an more gradual interface.  As will be 
seen in the full dimensional calculations, decreasing the chain 
stiffness allows the magnitude of the fluctuations to change more 
rapidly between successive monomers, as expected.

%========================================================================
\subsection{Fine Structure of the Free Energy Profile:
Multiple Transition States and Folding Routes}
%========================================================================
We follow the same analysis outlined in the two-dimensional illustration
to describe the folding paths calculated in the full variational space,
but the free energy profile is evaluated only at the 
saddle-points and local minima rather than as a continuous path.

In this study, we focus on how the folding routes depend
on the persistence length of the chain.
To put various parameters into context, the homopolymers
polyglycine, polyalanine and polyproline 
have persistence lengths%
\cite{fn:ptw00a_stiff}
$l =$ 6\AA\ $(\approx 2a)$, 20\AA $(\approx 5 a)$,
and 220\AA\ $(\approx 60 a)$, respectively.
\cite{flory:69,cantor_schimmel:80}.
In the freely rotating chain model $l = a/(1-g)$, so that
these persistence lengths correspond to the chain stiffness 
parameters $g = $ 0.5, 0.8, and 0.98, respectively.
Modeling the protein  backbone with a single uniform chain 
stiffness parameter, we take the chain stiffness of polyalanine 
to be a reasonable value for the protein backbone.

We report the free energy profile at the folding transition temperature,
$T_{{\rm f}}$. This is the temperature at which the folded and globule ensembles
have the same equilibrium probability 
$(\Delta F_{{\rm G}} = \Delta F_{{\rm N}})$.
Since the entropy and energy loss in these states depend on the 
chain stiffness, $T_{\mathrm{f}}$ is different for different persistence lengths.
(For a flexible chain $l = 2  a$, 
$k_B T_{{\rm f}}/\epsilon_0 \approx 1.2$, whereas
for a very stiff chain 
$l = 20 a$, $k_B T_{{\rm f}}/\epsilon_0 \approx 2.2$).
Similarly, the unconstrained ensemble 
(which we define as the zero of the free energy) is dependent
on the persistence length.
Consequently, to compare the folding profiles of chains with 
different persistence lengths, it is convenient 
to plot the free energy profile relative to the globule free energy 
against a normalized energy coordinate
\begin{equation}\label{eq:enorm}
{\bar E} = 
(\Delta E - \Delta E^{{\rm G}})/(\Delta E^{{\rm N}} - \Delta E^{{\rm G}}), 
\end{equation}
where $\Delta E^{{\rm G}}$ and $\Delta E^{{\rm N}}$ are the globule and 
native energy changes, respectively.
${\bar E}$ is the fractional stabilization energy and equals
0 at the globule state and 1 at the native state.

\begin{figure}%[h]
\psfig{file=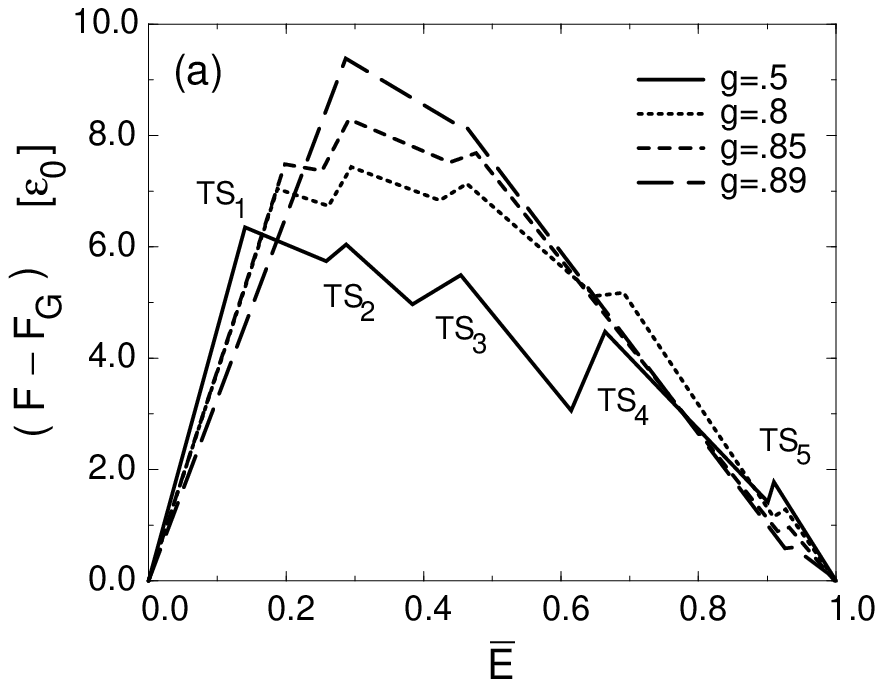,width=3.3in}
\psfig{file=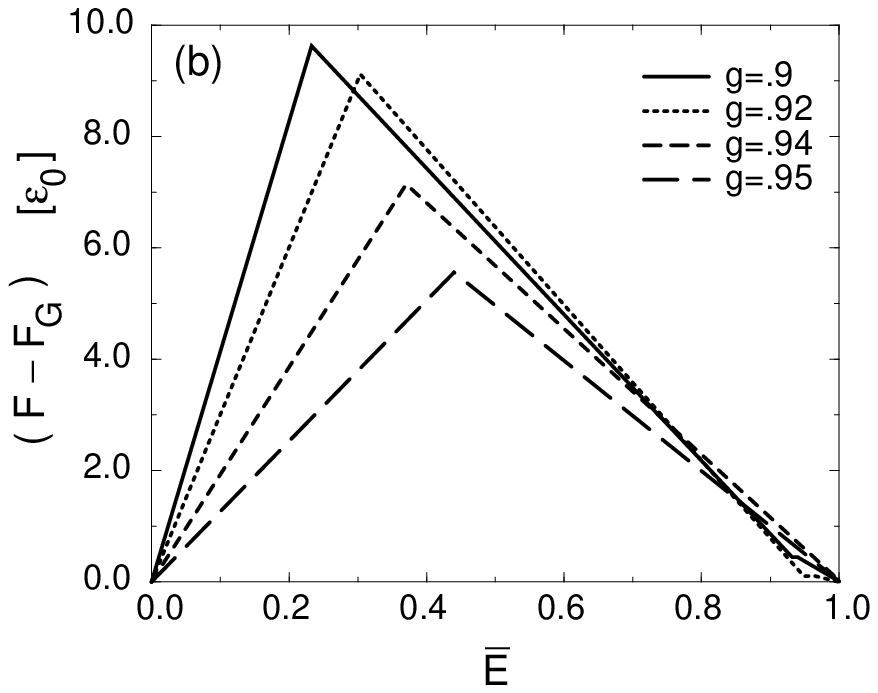,width=3.3in}
\caption{Free energy profile vs. normalized energy
[Eq.(\ref{eq:enorm}],
for different persistence lengths:
(a) $g = 0.5,.8,.85,.89$. (b) $g = 0.9,.92,.94,.95$.}
\label{fig:fprofile}
\end{figure}

Fig. \ref{fig:fprofile} shows the free energy profile for chain stiffness
parameters ranging from $g = 0.5$ to $g = 0.95$ 
($l = 2 a$ to $l = 20 a$).  
The free energy profile versus energy for a flexible 
chain ($l = 2 a$) is shown as the solid curve in Fig. \ref{fig:fprofile}a.
The profile exhibits five transition states
(saddle-points) separated by local minima.
We specify the transition states sequentially from the globule minimum
(G) to the native minimum (N),  and the minima by
the index of the adjacent transition state
(e.g., the minimum between ${\rm TS}_1$ and ${\rm TS}_2$ is 
denoted by ${\rm min}_{12}$). The profile can be described as
a rugged fine structure resulting from the different
structural ensembles of the local minima superimposed on 
a single average free energy barrier.  
The fine structure on the profile is modest in magnitude amounting
to a stabilization of a ``high energy intermediate'' by at most
1$k_B T_{\mathrm{f}}$.
This fine structure should 
not be confused with the ruggedness due to frustration \cite{jdb:pgw:87},
which we call ``transverse'' ruggedness, coming from degrees of freedom
different from the one plotted; instead, the fine structure is
better described as ``longitudinal''
ruggedness (along the reaction coordinate) \cite{ms:mo:97}.
This is a common feature of many models even those
which consider only native contacts \cite{vsp:dsr:99} and which therefore
have perfectly funnel-like surfaces.

Starting with the most flexible chains shown in Fig. \ref{fig:fprofile}a, as 
the stiffness increases the magnitude of the barrier increases and the 
the minima separating the 
transition states become relatively still more shallow 
(the longitudinal ruggedness diminishes).  
A flexible chain can take advantage of particularly
strong contacts while losing a relatively small amount of entropy to localize the
pair. This gives rise to the possibility of relatively stable local 
minima and lower free energy 
barriers. Stiffer chains must localize larger segments of the
chain (on the order of the persistence length) resulting fewer distinct
local minima and larger free energy barriers.  For the 
largest chain stiffness in Fig. \ref{fig:fprofile}a ($l \approx 9 a$)
the local minima have nearly disappeared, leaving a free energy profile
with a single transition state ensemble.  
The profile for even stiffer chains 
($l = 10 a$ to $l = 20 a$)  is shown in Fig. \ref{fig:fprofile}b.  
In this parameter range, as the chain stiffness increases further
the barrier decreases and the single transition state occurs at a 
larger fractional stabilization energy (i.e., 
the energy of transition state becomes closer to the native energy).  
When the persistence length is large enough, large sections of the 
chain must be constrained resulting in a more folded transition 
state.  Evidently, for very stiff chains
the transition state is similar enough to the native minimum that the
barrier decreases.

\begin{figure}%[h]
\psfig{file=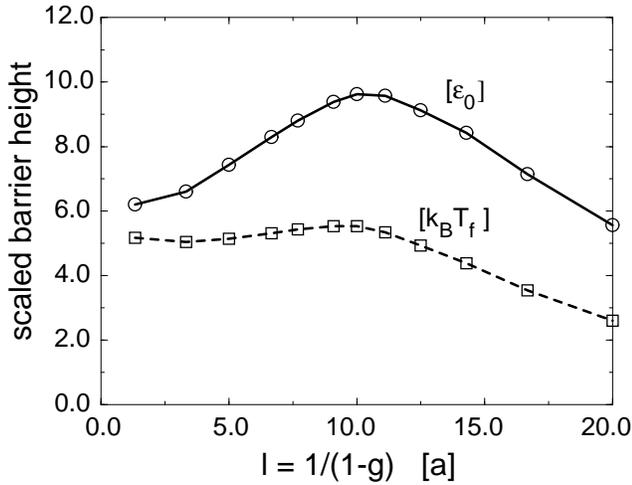,width=3.3in}
\caption{Barrier height vs. persistence length.  
Free Energy scaled by Miyazawa-Jernigan energy scale 
$\epsilon_0$ (solid), and by the folding temperatures 
$k_B T_{{\rm f}}$(dashed).
}
\label{fig:bh_vs_l}
\end{figure}

The barrier height is plotted as a function of persistence length in 
Fig. \ref{fig:bh_vs_l} (solid line).  The maximum 
barrier height occurs for a persistence length $l \approx 10 a$ with
an increase of approximately $70 \%$ relative to the barrier height
for the most flexible chain considered. However, the appropriate energy 
scale for the transition is $k_B T_{{\rm f}}$. In units of 
$k_B T_{{\rm f}}$, the barrier height is relatively constant 
for a wide range of persistence lengths (dashed line), changing only 
approximately $10 \%$ for persistence lengths up to $l \approx 12 a$.
The robustness of the barrier height at $T_{\mathrm{f}}$ is interesting
since the persistence length of real proteins in the laboratory
is not precisely known.

\begin{figure}%[h]
\psfig{file=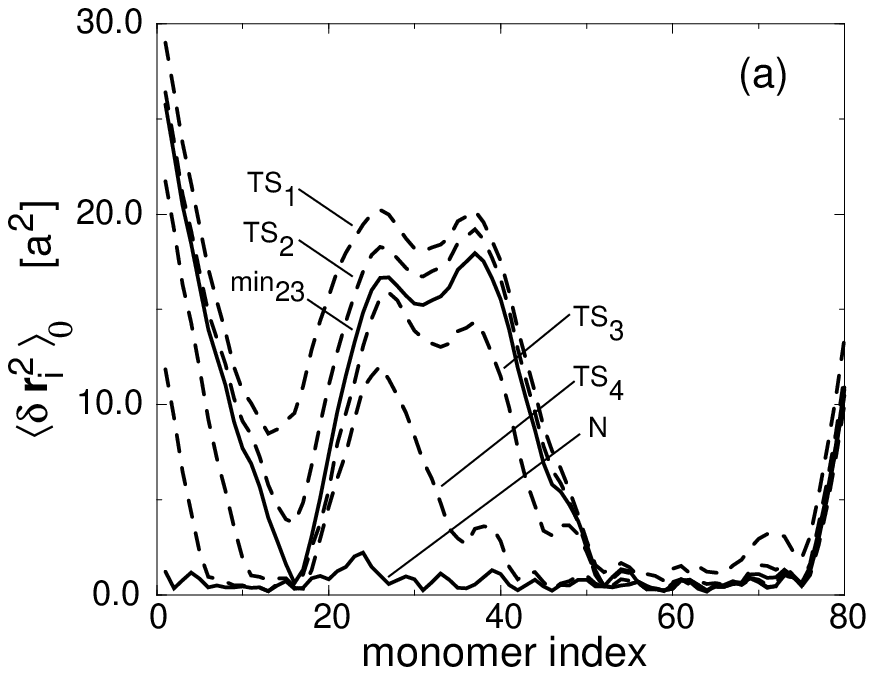,width=3.3in}
\psfig{file=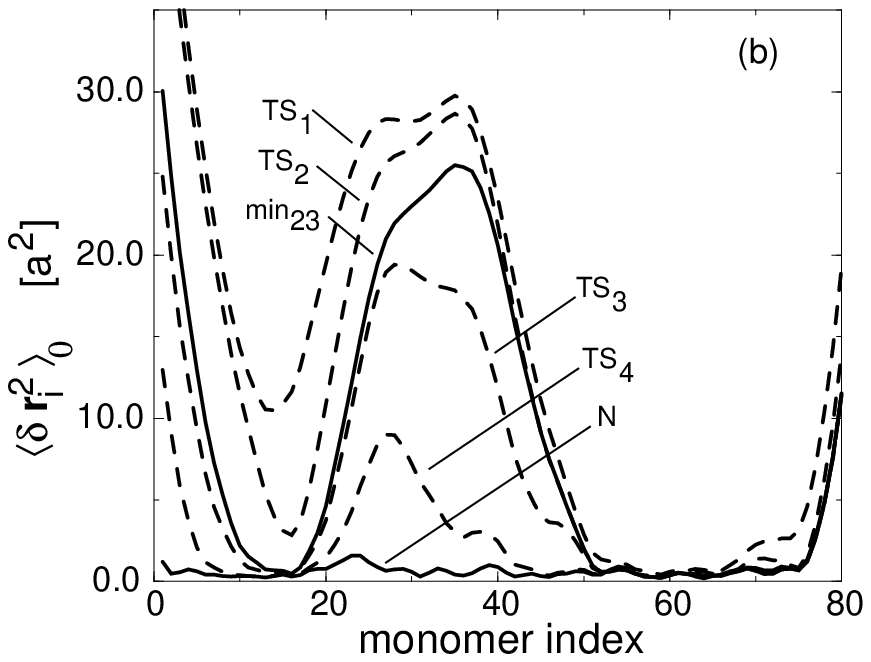,width=3.3in}
\psfig{file=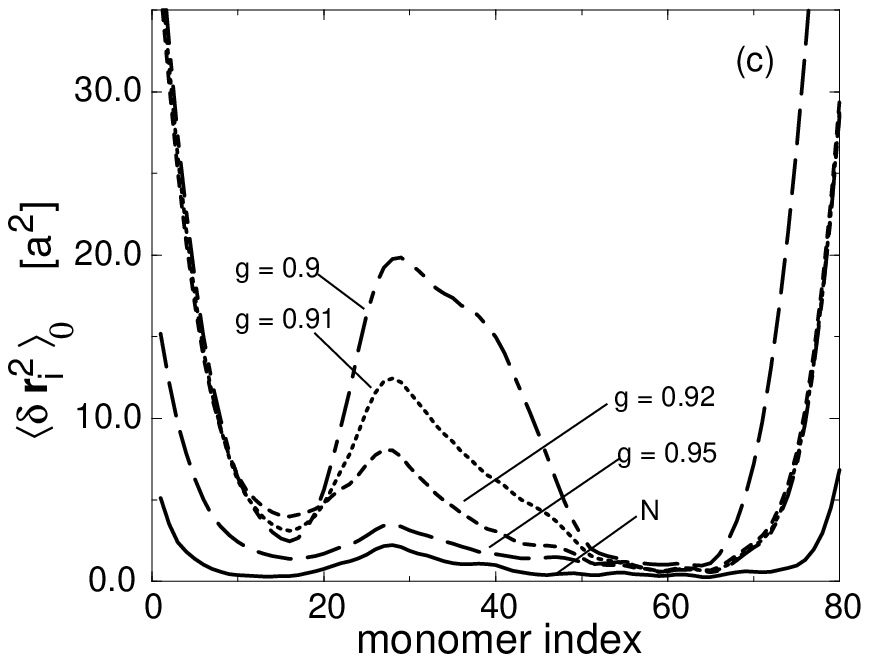,width=3.3in}
\caption{Fluctuations vs. sequence index,
$\langle \delta {\bf r}^2_i \rangle_0 = G_{ii} a$ 
where a is the distance between successive monomers, 
of selected stationary points on the folding route
for different persistence lengths. 
(a) $l = 2 a$ ($g = 0.5$), (b)  $l = 5a$ ($g = 0.8$), 
(c) $l \approx 10,11,13,20$ ($g = 0.9,0.91,0.92,0.95$).}
\label{fig:gii}
\end{figure}

Each ensemble of structures along the average folding route can 
be characterized by the local temperature factors (Eq.\ref{eq:tfac}).
The fluctuations corresponding to the transition states
for a flexible chain ($l = 2 a$) are shown as the dotted curves
in Fig. \ref{fig:gii}a.  The fluctuations corresponding to the local
minima are within those of adjacent saddle-points; for clarity, 
only one local minimum, ${\rm min}_{23}$, is shown (solid line).  
The folding route
can be characterized by considering the structure of the transition
states from the globule to the native structure.  The first
transition state ensemble ${\rm TS}_1$ is described by the ordering of 
helices H4--H5, stabilized by the partial localization of a region 
of helix H1, while residues in helices H2--H3 remain de-localized. 
In the subsequent transition states, helix H1 becomes progressively
more ordered, while helices H2--H3 continue to have large fluctuations
and are the last to order.  This scenario for the folding of 
$\lambda$-repressor agrees with the interpretation of kinetic 
data based on $\phi$-value analysis \cite{bhdco:97a}.

The temperature factors for a larger persistence length ($l = 5 a$)
are shown in Fig. \ref{fig:gii}b. The behavior of the 
fluctuations describing the
transition state ensembles is qualitatively similar to the 
more flexible chain, though the magnitude of the fluctuations of 
the disordered regions is larger.
Some of the detailed features shown in Fig. \ref{fig:gii}a have been
smoothed out since the chain stiffness suppresses variations
along the sequence smaller than the persistence length. For example,
the very specific localization of a helix H1 residue shown in
the ${\rm min}_{23}$ curve of the more flexible chain (Fig. \ref{fig:gii}a)
has been broadened to a larger region for the stiffer chain.  
These differences are rather subtle, but the comparison is useful
to illustrate the progression to larger chain stiffnesses.
The fluctuations for the single transition state characteristic of larger 
persistence lengths is shown in 
Fig. \ref{fig:gii}c.  (Note, here the different curves correspond to
different persistence lengths rather than transition states along
the same folding route.) While the general shape of the curve is
maintained, the magnitude of the fluctuations become increasingly more like 
those of the native state, and in the largest persistence length considered
$l = 20 a$ only the end segments of the chain are significantly
disordered.  

%======================================================================
\subsection{Contiguous Sequence Approximations}
%======================================================================
Several recent and apparently accurate estimates of folding kinetic 
parameters have assumed that the transition state ensemble can 
be described by assigning contiguous segments to be either folded
or not and allowing the sequence to be passed into such fully folded or 
unfolded configurations \cite{ovg:avf:99,ea:db:99,vm:wae:99}.  
As Plotkin et al. argued, such a contiguous sequence approximation 
(CSA) should apply to late transition states where the entropy 
is low \cite{ssp:jw:pgw:96a}.  How does the polymer's characteristics 
determine the quality of this approximation?

Following the contiguous sequence approach, to simplify the problem
we reduce the number of states in the variational theory's description
of the protein ensembles by restricting the
conformations to only those in which the structured residues are fully native
and  contiguous in sequence (single CSA), 
or alternatively, we consider two contiguous stretches of native
residues (double CSA)%
\cite{fn:ptw00a_csa}
.
For each fixed value of the number of folded residues, $N_{\mathrm{f}}$,
we find the minimum free energy configuration (satisfying the 
single or double contiguous  constraint).
In this way we can construct a free energy profile as a function of $N_{{\rm f}}$.  
This construction 
neglects the connectivity of the path since the minimum free energy
configuration with $N_{{\rm f}}$ folded residues
may not be simply related to that 
with $N_{{\rm f}} + 1$.  This connectivity
is an added complication to the approach and can be treated by 
the methods presented Refs. \citen{ea:db:99,ovg:avf:99}.  For the purposes of 
this illustration, we neglect this aspect of the approximation.
The approach outlined here is a great simplification of the more complete
variational formalism since it avoids the relatively difficult numerical calculation
of finding saddle-points as a function of degree of ordering.   
Still, the number of configurations needed in the double CSA
is quite large, but tractable 
($\approx 1.7$ million for $\lambda$-repressor).  
On the other hand, these approximations are rather restrictive 
since they neglect partial ordering and provide a less microscopic 
characterization of the folding route. 

One issue in comparing the exact variational and the CSA
results is how to measure the partial order characterizing 
the stationary ensembles in a global way to comparable the number
of fully native residues in the CSAs.
For a given stationary point, we can compare by estimating 
$N_{\mathrm{f}}$ through the normalized fluctuations
\begin{equation}\label{eq:tfac_norm}
\bar{{\cal B}}_i = 
({\cal B}_i - {\cal B}_i^{{\rm G}})/
({\cal B}_i^{{\rm N}} - {\cal B}_i^{{\rm G}}),
\end{equation}
where the superscripts G and N denote the fluctuations evaluated at
the globule and native state, respectively.  As a rough approximation,
we define $N_{\rm f}$ to be the number of residues with 
${\bar {\cal B}}_i \ge 0.95$.

\begin{figure}%[h]
\psfig{file=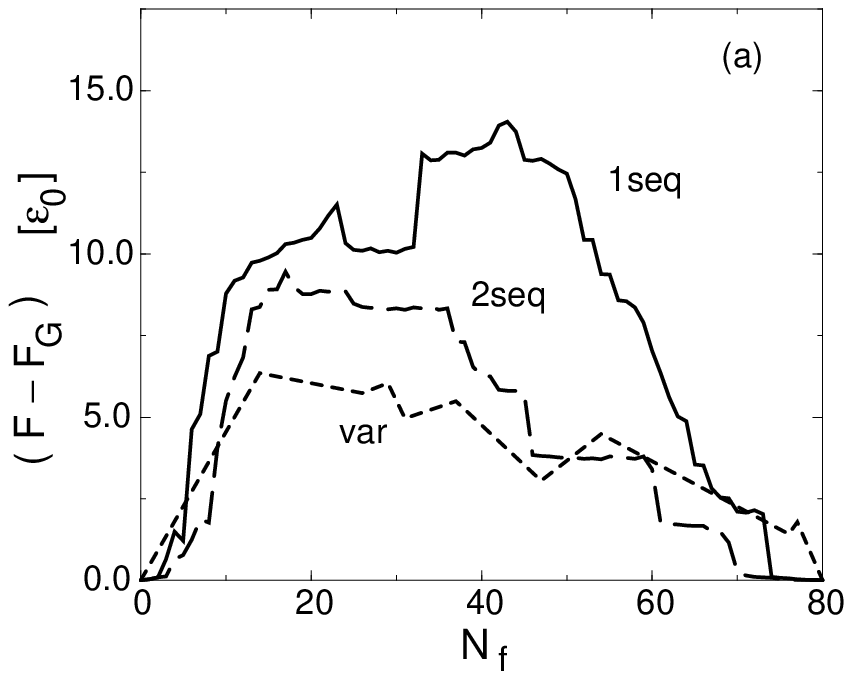,width=3.3in}
\psfig{file=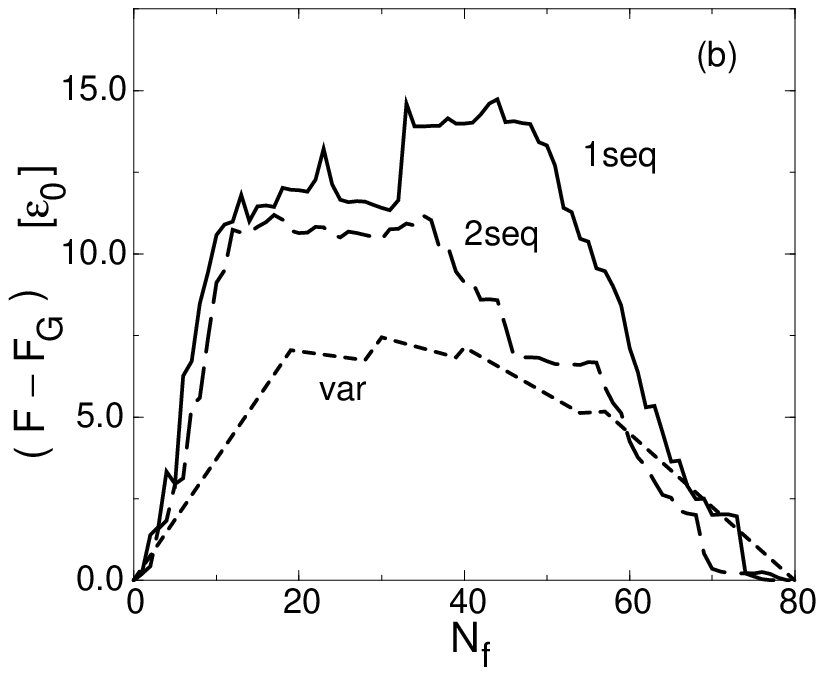,width=3.3in}
\psfig{file=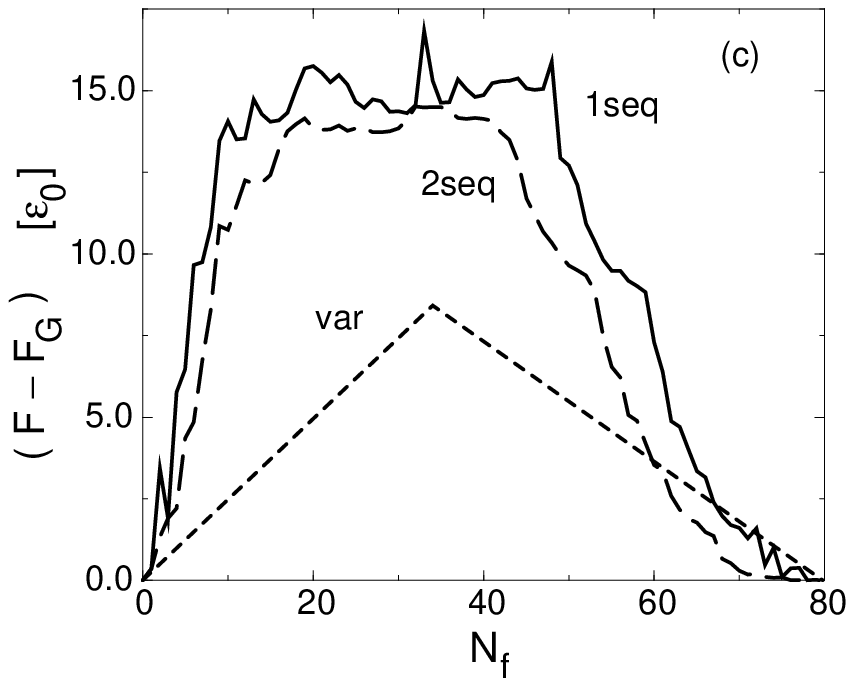,width=3.3in}
\caption{Free energy vs. number of folded residues for 
persistence lengths (a) $l = 2a$ ($g = 0.5$), (b) $l = 5a$ ($g = .8$),
and (c) $l \approx 14 a$ ($g = 0.93$) scaled by the Miyazawa-Jernigan 
contact energy scale, $\epsilon_0$.  The different curves correspond to 
contiguous sequence approximation (solid), double approximation (dashed),
and the variational theory (dotted).  For the variational profile,
$N_f$ is defined by the number of residues with
$\bar{{\cal B}}_i \ge 0.95$ (see text).
}
\label{fig:cntgf}
\end{figure}

The free energy profile from the variational theory (dashed),
the single CSA
(long-dashed), and double CSA (solid)
are plotted as a function of $N_{{\rm f}}$ for three different 
persistence lengths in Fig. \ref{fig:cntgf}.  
The barrier heights from the simpler approximations are about
twice the barriers 
obtained from the variational calculation for each persistence length.  
This is consistent with results of the other contiguous sequence approaches, 
where the barriers are much larger than obtained from
simulations of G\=o models with pairwise additive forces
\cite{hn:aeg:jno:98,js:jno:clb:99,mpe:pgw:00}.  
In the present comparison, one could also 
expect this behavior, since a variational Hamiltonian
with fewer degrees of freedom generally gives higher barriers 
(c.f. the barriers in the two-dimensional illustration 
Fig. \ref{fig:2d_fprofile} with the full dimensional calculation 
in Fig. \ref{fig:fprofile}).
Nevertheless, the simpler approximations provide an
intuitive explanation that agrees
qualitatively with the magnitudes of the barrier heights.  
Confining the residues to be either folded or unfolded (in contiguous
stretches) is responsible for the larger barrier heights, i.e.,
partial ordering reduces the barrier. This reduction seen in the 
more accurate calculation
is reminiscent of the way wetting between two stable phases in 
nucleated phase transformations lowers nucleation barriers by reducing 
surface tension, and is discussed in the context of protein folding 
in Ref. \citen{st:pgw:97b,pgw:97}.

Considering the free energy profiles for the two approximations 
for a flexible chain ($l = 2a$) plotted in Fig. \ref{fig:cntgf}a,
the maximum free energy for the single CSA
is approximately $50\%$ larger than the barrier from 
the double CSA and occurs at a greater value of 
$N_f$. Again there is no surprise in finding the barrier from the 
double contiguous approximation is lower, since the single contiguous 
configurations are a subset of the double contiguous configurations. 
It is interesting, however, to compare this difference as the
chain stiffness increases.
For a larger chain stiffness corresponding to 
polyalanine ($l = 5 a$) shown in Fig. \ref{fig:cntgf}b, the difference
in the barrier height from the single and double CSAs decreases, with the
single CSA barrier approximately $30\%$ larger than the double CSA barrier.
Increasing the stiffness further ($l \approx 14 a$), the
free energy profiles shown in Fig. \ref{fig:cntgf}c are more similar,
differing by only $10\%$.  
Since the profiles become still more similar
as the chain stiffness increases 
(even though the conformations in the double CSA are less restricted), 
this comparison suggests that the single CSA more accurately describes stiff
chain conformations than it does flexible chain conformations.

\begin{figure}%[h]
\onecolumn
\psfig{file=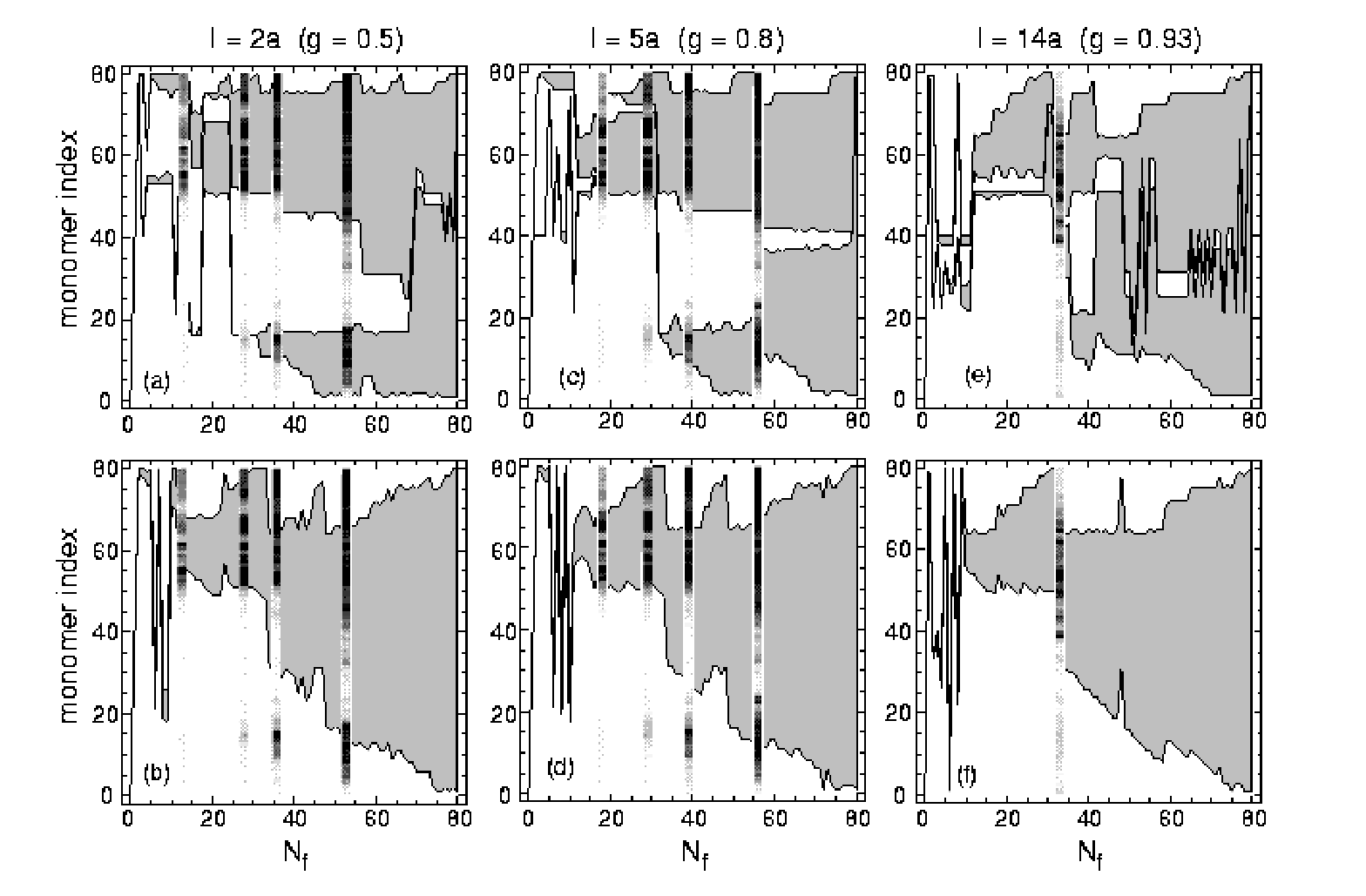,width=7.in}
\caption{
Configuration of native residues vs. number of folded resides
persistence lengths for the double (top: a, c, e) and the
contiguous (bottom: b, d, f) 
sequence approximations. 
The columns correspond to the persistence lengths: 
$l = 2a$ (a,b), $l = 5a$ (c,d), $l = 14a$ (e,f).
Residues set to native constraints are indicated by 
the shaded region.  The density plot corresponds to the normalized 
native density (Eq.\ref{eq:ndens_norm}) evaluated at the 
transition states for each persistence length.
(Black to white represents $\bar{\rho}_i =$ 0 to  1.)
The ordinate for the density plot, $N_f$, is  
defined by the number of 
residues with $\bar{{\cal B}}_i \ge 0.95$ (see text).
}
\label{fig:cntg_struct}
\end{figure}

To make this connection more precise, we consider 
which residues are ordered along the folding route in 
these approximations.  In the present treatment, this information can
be easily represented by a plot of the folded regions specified by
the monomer index as a function of $N_f$.  
The structured parts of the chain 
are indicated by the shaded regions in 
Fig. \ref{fig:cntg_struct}.
To characterize the structure at the saddle-points of the variational
free energy surface we consider the Gaussian measure to the 
native structure
\begin{eqnarray}\label{eq:ndens}
\rho_i &=& 
\left\langle
\exp \left[
-\frac{3}{2a^2} \alpha^{{\rm N}}({\mathbf r}_i - {\mathbf r}_i^{{\rm N}})^2
\right]
\right\rangle_0  
\nonumber \\
&=&
(1 + \alpha^{{\rm N}} G_{ii})^{-3/2} 
\exp 
\left[ - \frac{3}{2a^2} 
\frac{\alpha^{{\rm N}}({\bf s}_i - {\bf r}_i^{{\rm N}})^2 }
{1 + \alpha^{{\rm N}} G_{ii}} 
\right]
\end{eqnarray}
This measure of the monomer density relative to the native position
is the order parameter employed in the companion paper to study the 
dynamics of the barrier crossing.  The degree of native structure at
the transition state can be characterized by the normalized measure
\begin{equation}\label{eq:ndens_norm}
\bar{\rho}_i = ( \rho_i - \rho_i^{{\rm G}})/
(\rho_i^{{\rm N}}-\rho_i^{{\rm G}})
\end{equation}
where the superscripts G and N denote Eq.(\ref{eq:ndens}) evaluated
at the constraints corresponding to the globule and native states, 
respectively.

Consider first the folded residues from the
double CSA for a flexible chain ($l = 2 a$)
shown in Fig. \ref{fig:cntg_struct}a. Consistent with the
full variational results, the shaded region
clearly shows the structure forms between within helices H4--5 and helix
H1.
Superimposed on the regions of folded residues (from the double
CSA) are $\rho_i$ as function of monomer index for the four main 
transition states of the folding route of a chain of the same stiffness
(from the variational theory).
The density plots of $\rho_i$ indicate that the relatively unfolded
regions  are indeed more structured than
the globule value $\rho_i = 0$. In particular, at the interface
between the folded and unfolded regions in the double CSA,
$\rho_i$ obtains intermediate values indicating partial ordering.  
Nevertheless, the folded residues in the double CSA
agrees qualitatively with the structure obtained from the 
variational saddle-points.
In contrast, the folded residues in the 
single CSA do not agree with the variational theory very well at all, 
as shown in Fig. \ref{fig:cntg_struct}b.

Figs.\ref{fig:cntg_struct}c and \ref{fig:cntg_struct}d show
structured regions along the folding routes for a greater chain stiffness
$l = 5a$, the value appropriate for polyalanine.
For this chain stiffness, the unfolded region between helix H1 and 
helices H4--H5 in the double CSA closes at a smaller value of $N_f$ compared 
with the more flexible chain.  In this sense the structured residues from
the two CSAs are in better qualitative agreement,
though the discrepancy between the two is still pronounced.
This difference between the two approximations is greatly reduced for a 
chain with a much larger persistence length ($l \approx 14 a$) shown 
in Figs.\ref{fig:cntg_struct}e and \ref{fig:cntg_struct}f.  
While there are still unfolded regions
between folded regions in the double sequence approximation, they
only persist for a very limited range of $N_f$.  For this persistence length,
the variational free energy surface has only one transition state as
indicated by the density plot of $\rho_i$.  
The structure 
determined by the saddle-point agrees qualitatively with the structure
indicated by double CSA, but also with the single CSA 
(because the two are similar).

%**********************************************************************
\section{Conclusion}

In this paper, we used a variational approach to calculate 
the free energy profiles and characterize the transition state ensembles 
applicable when nonnative contacts can be ignored.  Using $\lambda$-repressor
as an illustrative example, we investigated the role of chain stiffness
on the fine structure of the free energy profile.  
We found that increasing the persistence length of the chain 
tends to smooth the free energy
profile, making longitudinal ruggedness less pronounced.  
The transition state ensemble with very stiff chains was found 
to be more folded than the ensemble with more flexible chains.  
These results can be interpreted in terms of the tension between
taking full advantage of strong local contacts while still respecting the 
bending rigidity of the chain.
We also found that while the
absolute barrier height has a pronounced maximum as a function of 
persistence length, the barrier scaled by the folding transition 
temperature $k_B T_{\mathrm{f}}$ 
was relatively robust over a wide range of persistence lengths.  

This study allowed us to investigate the applicability of simpler 
contiguous sequence approximations proposed recently.
Both the free energy profiles and the folded residues along the folding 
routes suggest that the single CSA more
accurately describes stiffer chains.  
Since the folded residues in the single CSA 
is roughly independent of the persistence length,
the single CSA sequence approximation corresponds to a chain 
that has a longer effective  persistence length than is appropriate
for most natural proteins. 
The double CSA
is able to capture the appropriate folded regions near the free energy barrier.
On the other hand, the neglect of partial ordering leads to 
over-estimates of the absolute barrier height for the $\lambda$-repressor
protein.  Nevertheless, these approximations should be good enough to
compute the perturbations of the activated free energy 
found in protein engineering experiments 
that probe the transition state ensemble ($\phi$-analysis) \cite{arf:am:ls:92}.

%=================================================
\section*{Acknowledgments}
This work has been supported by NIH Grant No. PHS 2 R01 GM44557.
%=================================================

%**********************************************************************
\appendix
\section{}
In this appendix, we outline a derivation of the 
expression for the entropy of the constrained chain
given in Eq.(\ref{eq:svar2}).

The partition function of the harmonically constrained chain
is given by 
\begin{eqnarray}
Z_0 &=& \exp\left[- A \sum_i C_i {\bf r}_i^{{\rm N}} \right]
\nonumber \\
&\times& {\rm Tr} \:\:
\exp \left[
 - A \sum_{ij} {\bf r}_i \cdot \Gamma^{(0)} \cdot {\bf r}_j
+ 2 A \sum_i C_i {\rm r}_i^{{\rm N}}\cdot {\bf r}_i
\right],
\end{eqnarray}
with $A = 3/2a^2$ and ${\rm Tr}$ denotes $\int \! \Pi_i d{\bf r}_i$. 
These integrals are easy to evaluate by completing squares 
\begin{eqnarray}
Z_0 &=& (\pi/A)^{3n/2} (\det G)^{3/2}
\nonumber \\
&\times&
\exp \left[ -A \sum_i C_i ({\bf r}_i^{{\rm N}})^2 
+ A \sum_{ij} C_i {\bf r}_i^{{\rm N}} \cdot G_{ij} 
\cdot C_j {\bf r}_j^{{\rm N}}
\right],
\end{eqnarray}
where $G = \Gamma^{{\rm (ch)}-1}$. 
We note that  Eq.(\ref{eq:rave}) can be used to express
the second term in the exponent as
\begin{equation}
A \sum_{ij} C_i {\bf r}_i^{{\rm N}} \cdot G_{ij} C_j {\bf r}_j^{{\rm N}}
= A \sum_i C_i {\bf r}_i^{{\rm N}} \cdot {\bf s}_i
\end{equation}
in terms of the average monomer positions, $\{{\bf s}_i\}$.  
From Eq.(\ref{eq:svar}), 
combining the partition function with
\begin{equation}
A \sum_i C_i \langle ({\bf r}_i - {\bf r}_i^{{\rm N}})^2\rangle_0
=
A \sum_i C_i (G_{ii}a^2 + {\bf s}_i^2 + 2{\bf s}_i\cdot{\bf r}_i^{{\rm N}} 
+ ({\bf r}_i^{{\rm N}})^2)
\end{equation}
gives the entropy of the constrained chain (ignoring a constant factor)
\begin{eqnarray}\label{eq:ap:svar}
S[\{C\}] = \frac{3}{2} \log \det G  &+& \frac{3}{2} \sum_i C_i G_{ii}
\nonumber \\
&+& A \sum_i C_i {\bf s}_i^2
- A \sum_i C_i {\bf r}_i^{{\rm N}} \cdot {\bf s}_i
\end{eqnarray}

This expression is easier to interpret after simplifying the
last two terms.  Inserting the identity, $G\Gamma^{(0)} = {\bf 1}$,
into the last term and using Eq.(\ref{eq:rave}) gives
\begin{eqnarray}
 A \sum_i C_i {\bf r}_i^{{\rm N}} \cdot {\bf s}_i
&=& A 
\sum_{ikj} C_i {\bf r}_i \cdot G_{ik} \Gamma^{(0)}_{kj} \cdot {\bf s}_j
\nonumber \\
&=& A \sum_{kj} {\bf s}_k \cdot \Gamma^{(0)}_{kj}\cdot {\bf s}_j.
\end{eqnarray}
In this form, the last two terms of Eq.(\ref{eq:ap:svar}) can be 
combined
\begin{equation}
- A  \sum_{kj} 
{\bf s}_k \cdot (\Gamma^{(0)}_{kj}- C_k\delta_{kj})\cdot{\bf s}_j
= -A \sum_{kj} 
{\bf s}_k \cdot \Gamma^{{\rm (ch)}}_{kj}\cdot {\bf s}_j,
\end{equation}
where we have used the definition of $\Gamma^{(0)}$ given in 
Eq.(\ref{eq:gam0}).  Thus, we have
\begin{equation}
S[\{C\}] = \frac{3}{2} \log \det G  + \frac{3}{2} \sum C_i G_{ii}
- \frac{3}{2a^2} 
\sum_{kj} {\bf s}_k \cdot \Gamma^{{\rm (ch)}}_{kj}\cdot {\bf s}_j,
\end{equation}
which is Eq.(\ref{eq:svar2}).
%*************************************************************************
%*************************************************************************

%*************************************************************************
%*************************************************************************

\newpage
 
%***************************************************************************
%***************************************************************************

\begin{thebibliography}{10}
\bibitem{fn:ptw00a_csa}
The configurations in the contiguous sequence approximations are specified by
  the constraint parameters $\{C_i\}$: if the $i^{\mathrm{th}}$ residue is
  fully folded (unfolded), $C_i$ is set to the value of the corresponding
  constraint at the native (globule) minimum.
\end{thebibliography}
\end{document}